\documentclass[useAMS,usenatbib]{mn2e}
\usepackage{graphics}
\usepackage{subfigure}
\usepackage{graphicx}
\usepackage{amssymb}

\title[Galaxy star formation in different environments]{Galaxy star formation in different environments}

\author[R. E. Gonz\'alez and N. D. Padilla]{R. E. Gonz\'alez$^{1}$\thanks{E-mail:
regonzar@astro.puc.cl (REG); npadilla@astro.puc.cl (NDP)} and N. D.
Padilla$^{1}$\\
$^{1}$Departamento de Astronom\'\i{}a y Astrof\'\i{}sica, Pontificia Universidad Cat\'olica de Chile, Santiago, Chile\\
}

\begin{document}

\date{Accepted --- . Received ---}

\pagerange{\pageref{firstpage}--\pageref{lastpage}} \pubyear{2008}

\maketitle

\label{firstpage}

\begin{abstract}
We use a semi-analytic model of galaxy formation to study signatures of large-scale modulations in the
star formation (SF) activity in galaxies.  In order to do this we carefully define local and global
estimators of the density around galaxies.  The former are
computed using a voronoi tessellation technique and the latter are parameterised by the normalised distance
to haloes and voids, in terms of the virial and void radii, respectively.
As a function of local density, galaxies show a strong modulation in their SF, a result that is in agreement with 
those from several authors. 
When taking subsamples of equal local density at different large-scale environments,  
we find relevant global effects whereby the fraction of red galaxies diminishes for galaxies in equal local
density environments farther away from clusters and closer to voids.  
In general, the semianalytic simulation is in good agreement with the available observational results, 
and offers the possibility to disentangle
many of the processes responsible for the variation of galaxy properties with the environment; we find that
the changes found in samples of galaxies with equal local environment but different distances to haloes
or voids come from the variations in the underlying mass function of dark-matter haloes. There
is an additional possible effect coming from the host dark-matter halo ages, indicating that halo assembly
also plays a small but significant role ($1.14\sigma$) in shaping the properties of 
galaxies, and in particular, hints at a possible
spatial correlation in halo/stellar mass ages.  
An interesting result comes from the analysis
of the coherence of flows in different large-scale environments of fixed local densities; the
neighbourhoods of massive haloes are characterised by lower coherences than control samples, except
for galaxies in filament-like regions, which show highly coherent motions. 
\end{abstract}


\section{Introduction}

The current understanding of galaxy formation and evolution has improved considerably in
recent times owing to new, controversial results such as the downsizing scenario \citep{cim},
and the detailed chemical composition of possible galactic building blocks \citep{kau,geis}.
The former has promoted the inclusion of important processes into models for galaxy evolution
that had either been ignored or treated in a simplified way, 
such as the presence of active galactic nuclei in the
centres of galaxies \citep{bow,cro,lagos}.
The latter controversy has fostered studies of satellites and central galaxies of similar masses, to discover
that the search for the building blocks of present-day galaxies 
is a problematic endeavour, since the formation times play
an important role in the characteristics of galaxies (e.g. Robertson et al., 2005, 
Lagos, Padilla \& Cora, 2009).
Hints to these results have become available only recently, from studies of the importance of assembly
on the clustering of dark-matter (DM) haloes (e.g. Croton et al., 2007; Gao \& White, 2006) and on shaping the population
of galaxies in haloes or groups \citep{zap}.

It is currently thought that the star formation activity in galaxies is a complicated process where several
factors are interlaced, such as the inter-stellar medium (ISM), the cooling of hot gas, the chemical
compositions of the different gas phases, feedback processes which return energy and metals into
the ISM, and dynamical effects such as mergers, winds, tidal interactions (see 
for instance Cole et al., 2000; Baugh, 2007).  Furthermore, a successful model of galaxy formation
also needs to deal with the problematic formation of the first stars which form in 
a metal free medium \citep{gao3}.  Semi-analytic models assume that the first generation
of stars is simply able to occur, and from then on
adopt simple, observationally or theoretically motivated models for these evolutionary
processes.  These models allow a study of the measurable effects these assumptions induce on the resulting galaxy
population.  In particular, since the galaxies modeled this way are embedded in a cosmological numerical
simulation, it is also possible to study 
the role of environment in the star formation (SF) process, either using local density estimators 
or studying samples of galaxies residing in specific structures.

Studies of environmental effects in cosmology can 
focus on changes of the SF activity in galaxies with the degree of interaction with neighbour galaxies via 
mergers, or other dynamical effects.  This approach has been applied both to observations and simulations.  For instance,
\citet{bal2} find a significantly higher SF in low local densities with respect to dense local environments; 
in order to obtain this result they 
estimate the local density using the projected area enclosed by the fifth nearest neighbour ($\Sigma_5$). 
Since young stellar populations are associated to bluer colours, a high SF would produce a population of
blue galaxies. The present-day galaxy population shows a clear bimodal colour distribution 
with a blue component of star-forming galaxies, and a red component populated by passive galaxies \citep{bal,omil}.
As the local density increases, only the relative amounts of galaxies in the red and blue parts of the
distribution vary, leaving the location of the peaks unchanged.  This particular feature makes it possible
to use a simple measure, the red galaxy fraction, to study the variations in the galaxy population
with local environment.

When extending to global density estimators given by the large-scale structures in the distribution
of galaxies, it becomes particularly difficult to detect or predict a variation of the SF activity
for fixed values of the immediately local environment of the sample.
For instance, \citet{bal} use two definitions of local density convolving the galaxy distribution with
gaussian kernels of $1.1$ and $5.5$Mpc to study variations in the relation between  SF and density; they
find some indications of a possible dependence on the smoothing scale.
Large-scale effects on the SF of galaxies were also studied using specific structures. In galaxy Voids,  
\citet{cec} found that at fixed local densities there is a clear
trend towards bluer, more star-forming galaxies in void walls with respect
to galaxies in the field; this
represents the first clear detection of a large-scale effect on the SF.
In filaments, \citet{por} studied this dependence using data from the
2 degree Field Galaxy Redshift Survey
\citep{col} 
suggesting that the triggering of 
SF bursts can occur when galaxies fall from filaments towards cluster centres.
Furthermore, \citet{van} studied galaxy clusters in the Sloan Digital Sky Survey
\citep{sto} and found that the morphological type of satellite galaxies
depends on both, the local density as shown by \citet{dres}, and 
also on the normalised distance to the cluster centre 
as was previously shown by \cite{gom}, indicating that part of the effects in either
case are due to the proximity of the cluster centre rather than to the local environment within which
it is embedded.

With the aim to find clues on the physical origin of the different relations between SF and environment,
we make use of a numerical cosmological simulation with semi-analytic galaxies to study the variations 
of the SF activity in different global
and local environments using observational parameters such as colours and red galaxy fractions.  We
will compare the outcome of these studies to previous results from the 2dFGRS
and the SDSS to determine the degree out to which
a $\Lambda$ Cold Dark Matter (LCDM) model is able to imprint into a galaxy population the observed
large-scale SF trends.
Furthermore, the use of a simulation will allow us to determine and use
parameters which would otherwise be difficult or impossible to obtain from 
observations, including the SF history of individual galaxies, their merger histories, 
precise density measurements, and host DM halo masses, and to analyse their possible
role on shaping the distributions of galaxy colours and their dependence on the local and global environments.

This paper is organised as follows.  Section 2 introduces the semi-analytic galaxies used in this paper and the
procedures used to define global and local density proxies.  Section 3 
contains the analysis of the contribution from large-scale effects
on this local density relation by measuring red galaxy fractions, star-forming fractions, 
and analysing variations in the galaxy
host halo masses and possible systematic biases in analyses performed in both, simulations and observations.  Section
4 discusses possible causes for the large-scale modulation of SF, and finally Section 5 summarises the results
obtained in this work.

\section{Data Samples}

We use $68801$ galaxies corresponding to the $z=0$ output of the \small{SAG} 
semi-analytic code \citep{lagos} which includes 
gas cooling, star formation, mechanical and chemical feedback from SN types Ia and II, and follows
the evolution of central supermassive black holes and the corresponding
AGN feedback.  The underlying numerical simulation corresponds to 
a LCDM model within a periodic cube of $60$h$^{-1}$Mpc a side with $256^3$ particles,
for a mass resolution of $1\times10^9$h$^{-1}M_{\odot}$ per particle.  The dark matter haloes in the simulation
are characterised by at least $20$ particles, and reach a maximum mass of $\simeq 5\times 10^{14}$h$^{-1}M_{\odot}$.
The semi-analytic galaxy catalogue includes magnitudes in 5 bands, SF histories, stellar masses of 
the disk and bulge components, DM host halo mass,
galaxy type\footnote{
This galaxy type should not be confused with a morphological or spectral type.  It indicates
whether the galaxy is the central galaxy 
of the DM halo (type $0$); a satellite within a DM substructure (type $1$); 
or a satellite without a DM substructure (type $2$).
}
plus a range of other underlying parameters 
that comprise a complete history of each individual galaxy.

In order to analyse the dependence of the SF activity on environment, we first construct samples of 
semi-analytic galaxies by applying cuts in two types of density; a ``global" density estimator related to large-scale 
landmarks in the distribution of matter in the simulation, and a ``local" density estimator related 
to the typical distance to the nearest neighbour galaxies. In the case of observational data,
the latter is thought to be more closely related to the effects arising from galaxy-galaxy
interactions; the \small{SAG} model does not include galaxy-galaxy interactions but does include halo-galaxy
effects, such as the processes new satellites undergo when entering a new DM halo (see Lagos,
Cora \& Padilla, 2008).

\subsection{Global Density Definition}

We parameterise the global densities in which individual objects are embedded by measuring their
distance to notorious landmarks in the simulation.  Possible choices for landmarks 
include galaxy clusters and voids, which correspond to rare fluctuations in the 
density field.  Filaments could also be selected as landmarks but in principle these structures are located at the
void walls in our numerical simulation (i.e. $0.8-1.2 r_{void}$) and we therefore do not use them in our analysis.
In order to illustrate the meaning of a global density parametrised this way, notice that 
galaxies located on a spherical shell centred in, for instance, a Cluster of galaxies
could in principle be embedded in a wide range of local densities, depending on whether they
are field galaxies or part of filaments or groups.
For a fixed global environment the local environment can differ significantly.

DM haloes in the numerical simulation follow quite approximately NFW profiles \citep{nfw}, where
haloes of similar concentrations can be scaled to a single profile using a scale radius, $r_{200}$, which
encloses an overdensity of $200$ times the critical density in the universe.
Once this scaling is applied, the spherical shells distant by $r/r_{200}$ from any halo centre are characterised
by similar overdensities.  This approximation is also valid for the full population of haloes which 
presents a narrow range of possible concentrations \citep{hus}.
As the latter becomes an even better approximation 
when the population of haloes is restricted to a narrow range of masses, we select as landmarks for
global density estimators haloes with $M>10^{13}$h$^{-1}M_{\odot}$, for a total of $70$ selected DM haloes.
We then proceed to label galaxies according to their
 distance, in terms of $r_{200}$, to the closest DM halo within this sample.
From now on, we divide galaxies in $4$ subsamples at different distances from halo centres,
which we will refer to as $R_{H1}$ to $R_{H4}$ (with limiting values at $r/r_{200}=0.0,1.5,5,9$ and $20$). 
The corresponding 
average DM density around galaxies in each subsample ranges from $\approx 200 \rho_C$ to $\approx \rho_C$, where
$\rho_C$ is the critical mass density.

A similar principle applies to voids where their density profiles can be scaled using the void radius 
$r_{void}$ \citep{pad}; the profiles approach the average density in the Universe at $r/r_{void} 
\approx 1.5$ \citep{pat}.  
The void identification algorithm we adopt corresponds to the one described in \citet{pad}, and consists 
of a search of underdense spheres of varying radii within the periodic
simulation box, satisfying $\delta=\frac{\rho-\left<\rho\right>}{\left<\rho\right>}<-0.9$.
In the simulation
box we identify a total of $70$ voids with $r_{void}>4$h$^{-1}$Mpc, each containing in average a total of $\simeq 180$
galaxies in the range $r/r_{void}=0.8-1.2$.  
We use these voids to make a second parameterisation of global densities for the semi-analytic galaxies using
$r/r_{void}$.
We define $4$ distance ranges, referred to as $R_{V1}$ to $R_{V4}$, 
delimited by the values $r/r_{void}=0,0.55,0.85$, $1.05$ and $1.4$;
the average overdensity in these samples ranges from 
$\approx 0.05 \rho_C$ to $\approx \rho_C$.

\begin{figure*}
\begin{minipage}{180mm}
  \centering
  \vspace{0pt}
  \includegraphics[angle=0,width=0.8\columnwidth]{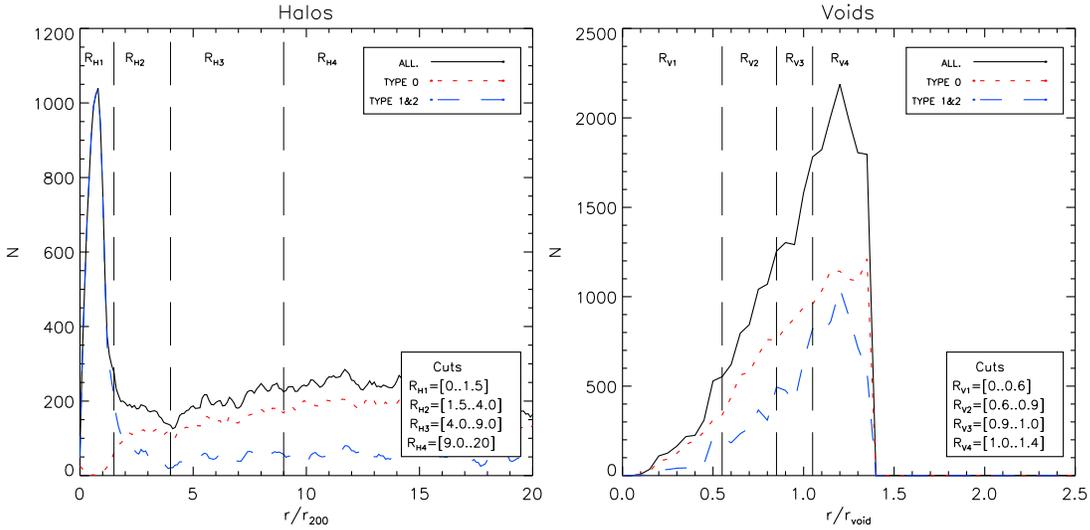}
  \caption{Distributions of normalised distances to the 70 largest DM haloes with $M > 10^{13}$h$^{-1} M_{\odot}$ (left
panel), and the 70 largest voids with $r_{void} > 4 h^{-1} Mpc$ (right panel).  Line types are defined in the figure
key.  The vertical long-dashed lines show the limits between the different global density samples defined in the text.}
  \label{fig:fig1}
\end{minipage}
\end{figure*}

The resulting distributions of normalised distances to haloes and voids for the semi-analytic 
galaxies in the simulation are shown in 
Figure \ref{fig:fig1}; the vertical long-dashed lines indicate the limits between different
global density samples selected according to the distance to haloes (left panel) and voids (right) in
the simulation.  Solid lines show the results for the full sample of galaxies in the simulation, 
dotted lines show central galaxies (notice the lack of objects near the halo centres, indicating the
typical minimum distance between haloes in the simulation), and dashed lines to satellite galaxies.  Except
for the lack of central galaxies near halo centres, the shapes of these distributions do not
change significantly with the galaxy type.

\subsection{Local Density Definition}

The problem of defining a measure of the local density around galaxies has multiple possible solutions, depending on the
relevant quantities that are to be associated to this estimate.  On the one hand
the standard approach of estimating a density using a fixed volume around a galaxy ensures
a fixed scale, but the extremely wide
dynamic range of densities ($\approx 8$ orders of magnitude) has the drawback of producing
low signal to noise estimates for low densities (dominated by Poisson noise), and oversampled
measurements at high density values.  Several works apply this method 
either by using a gaussian kernel to smooth the density distribution 
(e.g. Balogh et al., 2004a), or a more simple top-hat kernel of fixed size
\citep{cec}.
A choice of local density estimate associated to galaxy-galaxy interactions is the one taking
into account the closest neighbors of a galaxy by using the distance to the $N^{th}$ nearest neighbor.  The advantages
from using such an estimator relies in the likely relation between galaxy interactions and the SF in galaxies.
Most observational
studies adopting this latter approach use
adaptive projected 2D density estimators such as $\Sigma_5$; for instance, \citet{bal} compute $\Sigma_5$ 
using the distance 
to the fifth nearest neighbour brighter than $M_r=-20$ confined to a redshift slice of $\pm 1000km s^{-1}$ to
avoid biases from the finger-of-god effect. 

The alternatives to these approaches, generally applied to numerical simulations with
full 3-dimensional information, consist on
using an adaptive smoothing length 
proportional to the local particle or galaxy separation, or the galaxy-galaxy distance
(for Smoothed Particle Hydrodynamics, SPH), or 
adaptive local density estimates given by Voronoi Tessellations (VT, Voronoi, 1908).
With VT, each particle is associated to a domain volume so that every point inside this volume is
closer to the particle at its centre
than to any other particle; smoothing these density estimates with 
those of their immediate neighbors can be used to obtain a reliable 
measure of the local density.  This particular measurement method 
 called the Interpolated Voronoi Density (IVD), 
can be very useful for identification of bound objects \citep{platen,aragon,ney,rob}
in SPH simulations and has been shown to have a better resolution than any other adaptive density estimate including
SPH kernel smoothing techniques \citep{schaap,pel}. In the remainder of this work we adopt IVDs
for our estimates of local densities in the numerical simulation.

\begin{figure*}
\begin{minipage}{180mm}
  \centering
  \vspace{0pt}
  \includegraphics[angle=0,width=0.8\columnwidth]{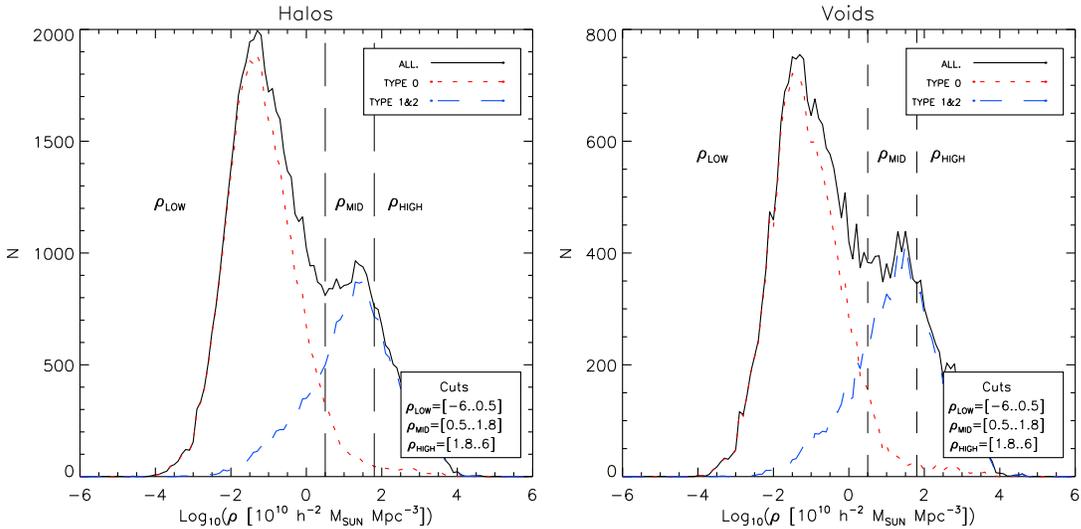}
  \caption{Distributions of IVDs for galaxies out to $20\times r_{200}$ from halo centres(left), and galaxies out to $1.4r_{void}$ from void centres.
Line types are defined in the figure key.  The vertical long-dashed lines show the limits between the different
local density samples.}
  \label{fig:fig2}
\end{minipage}
\end{figure*}

The left panel 
of Figure \ref{fig:fig2} 
shows the IVD distributions of galaxies out to $20r_{200}$ from the cluster centres (same as in the left
panel of {Figure} 1); the right panel shows the IVD distributions 
of galaxies out to  $1.4r_{void}$ from the void centres (as in the
right panel of {Figure} 1).
Regardless of whether the haloes (left panel) or voids (right panel) are used as global density landmarks, 
the distributions of local densities are similar since both selection criteria cover a large fraction
of the volume of the simulation.
The IVD distributions show a clear bimodal behaviour, where the peak at high densities is dominated
by satellite galaxies in massive DM haloes (long-dashed lines), and the density distribution
of central galaxies (short-dashed) reflects the density field around their host haloes, characterised by 
masses $M\gtrsim 10^{11}$h$^{-1}M_{\odot}$.

We use these density estimates 
to characterise the local environment of the semi-analytic galaxies.  In particular,
we will separate galaxies in three local density bins, to be referred to as the $\rho_{LOW}$, $\rho_{MID}$ and 
$\rho_{HIGH}$ samples.
The first tentative  cuts in stellar densities are applied at
 $3.16 \times 10^{10}$h$^{-2}M_{\odot} $Mpc$^{-3}$ and
$6.31 \times 10^{11}$h$^{-2}M_{\odot} $Mpc$^{-3}$,
selected so as to have a significant number of galaxies in each subsample.  Further restrictions in density
may be needed in order to ensure a constant median IVD in each sample studied.
For reference, the baryon density in the simulation is $\rho_b = f_b \times 2.8 \times 10^{11}$h$^{-2}M_{\odot} $Mpc$^{-3}$, 
with a baryon fraction $f_b=0.037$.

\begin{figure*}
\begin{minipage}{180mm}
  \centering
  \vspace{0pt}
  \includegraphics[angle=0,width=0.8\columnwidth]{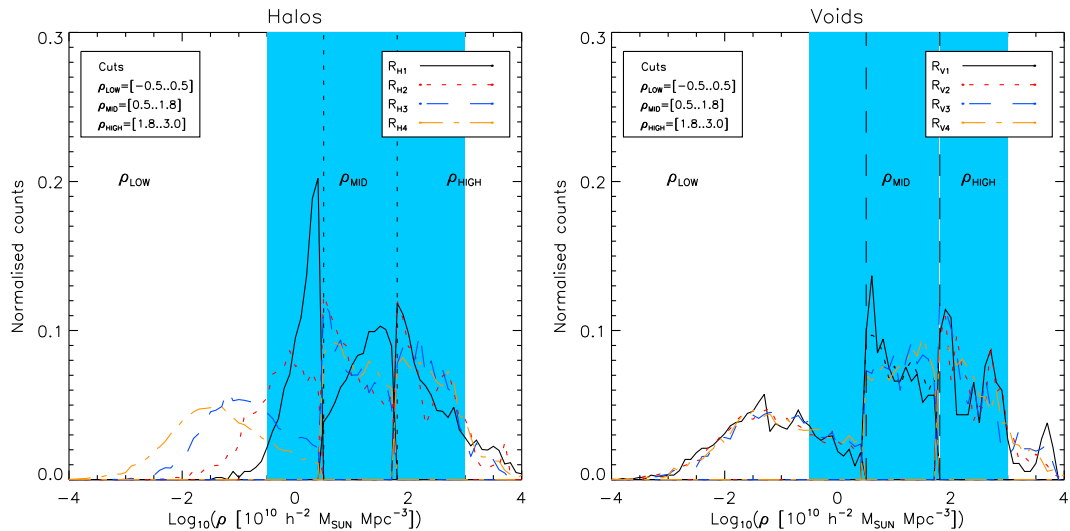}
  \caption{
IVD distributions independently normalised within ranges of IVD delimited by the vertical dashed lines, to
allow a comparison between samples corresponding to different large-scale environments (different line-types,
indicated in the figure key).  The shaded regions indicate the final local density cuts adopted so that different 
large-scale environments are characterised by roughly equal median IVD values.
}
  \label{fig:fig2b}
\end{minipage}
\end{figure*}

In the next section, we will compare samples of equal local density at different global environments.
It will become important to ensure that the median IVD values for different global environments 
are similar (if not, a measured 
global dependence could include some effects from the local density-morphology relation).  
In order to do this we apply further restrictions to the limits in local density.
Figure \ref{fig:fig2b} shows the IVD distributions renormalised independently 
within each preliminary local density cut (defined above).
As can be seen, the low and high local 
density cuts at different distances from halo centres (left panel) 
show very different distribution shapes and therefore are characterised by different
mean and median values for different large-scale environments;
adopting these preliminary cuts would induce a bias in the variation of galaxy properties with large-scale environment.
This problem is alleviated by including minimum and maximum IVD cuts for the 
low and high local density subsamples,
which will be defined from this point on using as limits,
$3.16 \times 10^{9}$h$^{-2}M_{\odot} $Mpc$^{-3}$,
$3.16 \times 10^{10}$h$^{-2}M_{\odot} $Mpc$^{-3}$,
$6.31 \times 10^{11}$h$^{-2}M_{\odot} $Mpc$^{-3}$, and 
$1.0 \times 10^{13}$h$^{-2}M_{\odot} $Mpc$^{-3}$.  
This ensures that the medians of the IVD distributions at different large-scale densities differ by less
than $0.15$ $dex$. 
In the following section we will extend our discussion on 
the effects of a varying median local density when comparing different global environments.

\section{Local and global effects on the SF}

In this section we will study colours and SF rates of model galaxies to study
their variations as a function of the local environment, and additionally to find out whether
the properties of galaxies located in similar local
environments show variations with the distance to voids and massive DM haloes (the global environment).
We will search for possible biases affecting the results from
simulations and  previous observational results.  
We will also study the variations
of the host DM halo mass function with the environment as an attempt to understand the changes in
the galaxy population.  We will finish the section by analysing the effect of the local and global
environment on the
fraction of star-forming galaxies.

\subsection{Galaxy colours}

The SF activity has a direct albeit complex
influence on the colours of galaxies, and therefore a modulation in either of these
quantities will be correlated with a modulation in the other, to some degree.  
Given that galaxy colours are more easily measured and
have been extensively used in environmental studies, we choose the $u-r$ colour to perform the following analyses.

We start by comparing the distributions of $u-r$ colours from \small{SAG} and SDSS. A useful
way to separate the two components of the bimodal colour distribution, consists on analysing galaxies
in different luminosity bins.  Bright galaxies are preferentially located at the red peak and faint galaxies tend to 
occupy the bluer peak of the colour distribution.   We define bright galaxies as those with z-band absolute
magnitudes $M_z<-21$, and faint
galaxies those with $M_z>-20$.
The colour distributions of bright and faint galaxies have been independently normalised to their integrals over colour.

\begin{figure*}
\begin{minipage}{180mm}
  \centering
  \vspace{0pt}
  \includegraphics[angle=0,width=0.8\columnwidth]{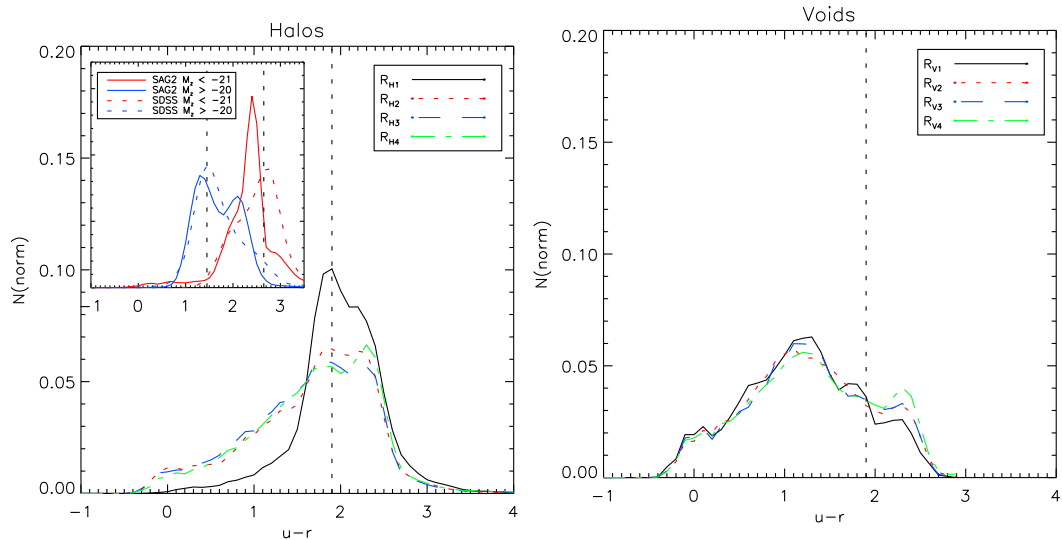}
  \caption{u-r colour distributions for different large-scale environments (indicated in the figure key) 
defined with respect to 
halo (left panel) and void (right panel) centres, using the normalised distance cuts shown in Fig. 1.
The inset in the left panel shows a comparison between SDSS and \small{SAG} $u-r$ galaxy colours (dashed and solid
lines, respectively), for faint and bright galaxies (blue and red lines, respectively).
}
  \label{fig:fig3}
\end{minipage}
\end{figure*}

The inset in the left panel of Figure \ref{fig:fig3} shows the $u-r$ colour distributions for
\small{SAG} galaxies with stellar 
masses higher than $10^8 M_{\odot}$ (solid lines), and SDSS galaxies(dashed lines). 
The red lines correspond to bright galaxies 
and the blue lines correspond to faint galaxies.
As can be seen the colour distribution from the SDSS is bimodal with peaks at roughly $u-r=1.45$
and $u-r=2.65$ (marked as vertical dotted lines), consistent with previous
observational results
\citep{bal2,pat}. The colour distribution of model galaxies is very similar to that of the SDSS, except for the 
small central peak that appears in the blue component of the distribution of semi-analytic galaxies, and for
a small shift in the red peak with respect to SDSS.   
Fortunately, these small differences do not affect 
the fractions of red galaxies (the focus of our study) in a significant way
since in both cases the colour distributions have clear dips at 
$u-r=1.9$, and the width of the blue and red components are very similar.
From this point on, we separate the blue and red populations with a colour cut at this value.
 
The main panels of {Figure} \ref{fig:fig3} show 
the $u-r$ colour distributions for samples in different large-scale environments (parametrised using the 
distance to haloes on the
left panel, and voids on the right) for the full \small{SAG} galaxy sample.
The lack of clear bimodalities in these distributions originates from the low luminosity/stellar
mass galaxies now included in the analysis.
As can be seen, the position of the colour peaks does not change with environment,
as was mentioned above, which allows the use of
red galaxy fractions as appropriate
measures of variations in galaxy colours as a function of local and global densities.
Fixed peak positions may be indicators that the transformation from blue to red colours would
either occur over a very short time-scale, or at high redshifts.  In the model, the rapid transition could
occur after the onset of AGN activity in a star-forming galaxy, on the one hand, but it is also possible that
the colour peaks are defined early on during the peak merger activity at redshifts $z>2$ (See for instance,
\citealt{lagos}).  

The left panel of
Figure \ref{fig:fig3} shows that the $R_{H1}$ subsample is clearly different than
the other samples defined using distances to haloes; it is clear that close to DM haloes most
of the galaxy colours populate the peak at
$u-r=2$.
Galaxies at larger distances from halo centres
show only a mild variation towards bluer colours. 
It can be noticed that galaxies around voids (right panel) also show a mild trend in the same direction,
of a redder galaxy population towards higher large-scale densities (the $R_{V4}$ subsample shows a higher
peak at $u-r=2.5$ than the other $R_V$ subsamples).

\subsection{Red galaxy fractions}

\begin{figure*}
\begin{minipage}{180mm}
  \centering
  \vspace{0pt}
  \includegraphics[angle=0,width=0.8\columnwidth]{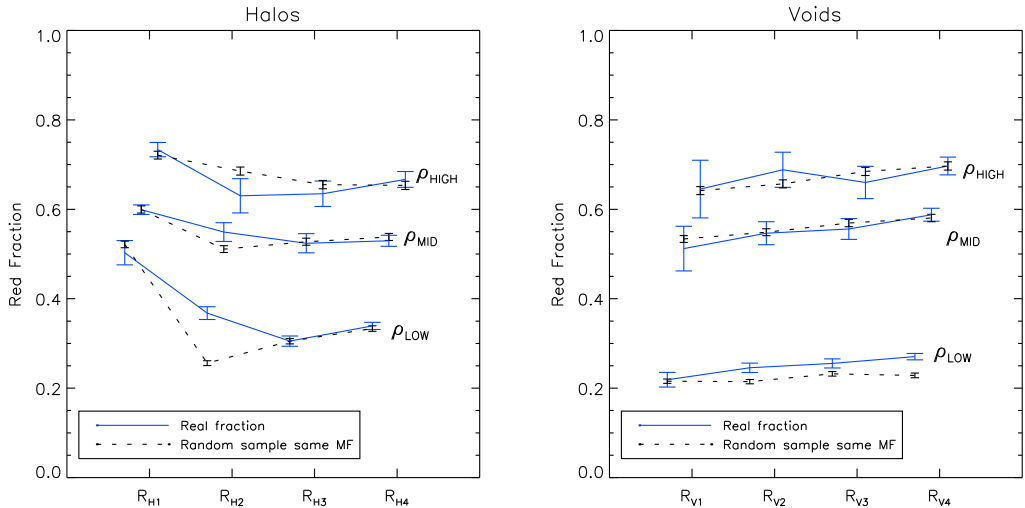}
  \caption{Red galaxy fractions as a function of distance to haloes (left) and voids(right panel) for
different local density cuts (blue solid lines).  The black dashed lines show the red galaxy
fractions for galaxies selected at random from the simulation so as to match the DM halo mass function and
the local densities of the parent subsamples.}
  \label{fig:fig4}
\end{minipage}
\end{figure*}

We compute red galaxy fractions using samples of galaxies located at different halo/void centric
distances at fixed local densities (the different samples are defined in the previous subsection, cf.
Figures \ref{fig:fig1} and \ref{fig:fig2b}).
We show the results in Figure \ref{fig:fig4}, where the solid lines show the variation of the red fractions
for subsamples at increasingly larger distances from
halo (left) and void (right) centres.  From bottom to top, the lines
correspond to bins of increasing local density; each line connects the results for samples with equal median IVD.
The errorbars are calculated using the jacknife method, which has been shown to provide robust estimates
of errors comparable to what is expected from sample variance (Padilla, Ceccarelli \& Lambas, 2005; all errors quoted in 
the remainder of this work are computed using this technique).  As the figure shows,
there is a clear dependence of the abundance of red galaxies on the large-scale environment,
with different significance levels depending
on the local density cut.  In particular, the effect of a changing red galaxy fraction is weaker 
as a function of the distance to the void centres
(it is significant at a $2.89\sigma$ level\footnote{
From this point on we will assess the statistical significance of our results using 
$n-\sigma$, where $n$ corresponds to the ratio of the difference
between two measured quantities and the standard deviation calculated using the errors of the 
mean added in quadrature (a student's test method).  }
between $R_{V1}$ and $R_{V2}$ for the lowest local densities).

Before describing in detail these effects, we 
construct new galaxy samples based on a selection of  
host halo masses and IVDs.  This will help the interpretation of the variations in the red fractions with the global
environment.
We measure the mass function of the host haloes of the galaxies in each sample selected according
to both, local and global density (notice that
as we include satellite galaxies, some of the haloes will be repeated) and produce new samples of galaxies
taken at random from the simulation box so that the mass functions and local IVD densities of the original samples
are reproduced.  Notice that these new samples do not include restrictions on their global environment.
The measured mass functions are shown in Figure \ref{fig:fig5}, and will be discussed in more detail later in this section.
We use these random galaxy subsamples to determine what changes are to be expected in the
red galaxy fraction from variations in the local density and underlying mass functions.  We make the comparison between
random and original samples in Figure \ref{fig:fig4} (dashed and solid lines, respectively).  
As can be seen, by comparing the adjacent dashed and solid lines,
most of the large-scale modulation of red galaxy fractions (solid lines) is produced by variations in the underlying
mass functions, as was suggested by \citet{cec} in their detection of this effect around
voids in the SDSS.  However, in particular for the low local density samples (lowermost sets of lines), there are some
effects that cannot be completely reproduced by the random samples with matching masses and local IVD densities.

The detailed results from 
Figure \ref{fig:fig4} can be summarised as follows,
\begin{itemize}
\item 
For any fixed local density there is always a trend of redder galaxies towards higher large-scale density
environments (i.e. close to massive DM haloes and further away from voids. The significance
of these trends ranges from $2.81\sigma$ to $8.58\sigma$ between $R_{H1}$ and $R_{H4}$ for the highest and lowest local 
densities respectively, and from $0.77\sigma$ to $3.45\sigma$ between $R_{V1}$ and $R_{V4}$ for the 
highest and lowest local densities respectively). In particular this trend, although weaker as a function of 
void-centric distance, is qualitatively consistent with the results from \citet{cec}. 
The largest effects around voids are found for the $\rho_{MID}$ local density range.
\item 
As a function of the distance to the halo centre,
changes in the red fraction are stronger at lower($8.58\sigma$ significance 
between $R_{H1}$ and $R_{H4}$) and intermediate($4.24\sigma$) local densities, with the strongest effects
found for galaxies near haloes and embedded in very low local densities.  It should be borne in mind that
the latter result could be influenced by the extreme treatment of satellites in the model, a subject we turn to
in the following Subsection.
\item At fixed global densities, changes in the red fraction as a function of the IVD are stronger in the outskirts 
of haloes and continue to change smoothly towards voids.  For instance, the curves corresponding to $\rho_{HIGH}$, $\rho_{MID}$ and $\rho_{LOW}$  
are spaced by $5.97\sigma$ between $\rho_{HIGH}$ and $\rho_{MID}$, and $3.29\sigma$ between 
$\rho_{MID}$ and $\rho_{LOW}$ near haloes ($R_{H1}$); and this separation
increases to $6.37\sigma$ between $\rho_{HIGH}$ and $\rho_{MID}$, and $8.94\sigma$ between 
$\rho_{MID}$ and $\rho_{LOW}$ away from haloes ($R_{H4}$). 
The fact that in haloes this dependence is weaker towards halo centers is in agreement with 
Gomet et al. (2003) who find that at local densities corresponding to virialised regions, the density-SF relation
becomes almost completely flat.
Towards lower global densities(in and around voids, from $R_{V4}$ to $R_{V1}$), the red fractions 
corresponding to low, mid and high local densities show the
largest variations, and also appear to converge to fixed ratios(with red fractions of $0.22$, $0.5$ \& $0.65$).  There is also a continuous transition starting at the centres
of haloes and going all the way to the inner void regions with the strongest variations in the 
transition from $R_{H1}$ and $R_{H3}$.  For instance, for the low local density case there is a $9.57\sigma$ detection
of a variation; this reduces to a $6.61\sigma$ change from $R_{H3}$ to $R_{V1}$).
\item 
From the analysis of differences between the red galaxy fractions for 
samples of galaxies at different global densities (solid
lines) and for galaxies selected so as to have the same host 
halo mass and IVD distributions (dashed lines), it can 
be seen that the large-scale modulation
is mostly the result of a variation of galaxy properties with their
host halo masses. Possible exceptions are the samples,
$R_{H2}$ for $\rho_{LOW}$, and $R_{H2}$ for $\rho_{MID}$, which show lower
measured red fractions ($5.88$ and $1.79\sigma$ lower, respectively) for the random/same mass function case, indicating an excess of red, low star formation 
galaxies towards cluster centres beyond any local density or mass function dependence. On 
the other hand, in $R_{H2}$ for $\rho_{HIGH}$ occurs the exact opposite. 
In voids, we find lower measured red fractions for the random/same mass function 
case, at $R_{V2}$, $R_{V3}$ \& $R_{V4}$ for $\rho_{LOW}$(with significances $2.65\sigma$, $2.04\sigma$ and $4.78\sigma$, respectively), indicating a diminished 
population of blue galaxies around voids.
In these cases there is 
a residual effect from another galaxy property which would correlate with 
the large-scale distribution of matter, which as can be noticed affects particularly the low or intermediate
IVD range.
\end{itemize}

\subsubsection{Tests on the reliability of our measurements}

The first bias we study is associated to a reported problem in semi-analytic models where satellite
galaxy colours (either in groups or clusters) are redder than is observed (e.g. Zapata et al., 2009).
Results from a recent model by \citet{font} suggest that the environment in groups and clusters is
less aggressive than previously assumed, since stripping processes may allow satellites to retain some of their
gas for longer periods of time; when adopting a mechanism that slowly removes the hot gas, the authors
are able to reproduce the observed satellite colours.  

In \small{SAG} satellites are completely stripped off their
hot gas after being incorporated to a halo, 
and consequently the resulting colours are approximately $0.1$ magnitudes redder than it is observed.
Therefore, in order to test whether this artificial reddening may affect our results
we repeat the analysis on the variations of the red fractions 
using only central isolated galaxies which would be less affected by the treatment
of gas in satellites.  

\begin{figure}
  \vspace{0pt}
  \includegraphics[angle=0,width=0.9\columnwidth]{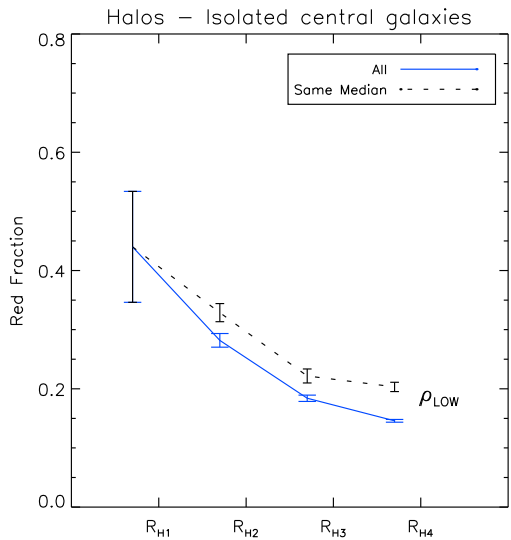}
  \caption{Red galaxy fractions as a function of the distance to massive DM haloes for
isolated central galaxies in low local density environments (blue solid line).  The black
dashed line shows galaxies in the $\rho_{LOW}$ sample with an additional cut to ensure fixed
mean and median IVD in the four subsamples at different distances to massive haloes.}
 \label{fig:figiso}
\end{figure}

Figure \ref{fig:figiso} shows the red fractions around haloes for isolated central galaxies 
located in  low local density environments (equal median values, dashed lines). 
As can be seen, the variations shown by the red fractions with global density are still very strong, 
with a similar amplitude compared to the results obtained using all the galaxies (cf. Figure \ref{fig:fig4}), but with
a lower abundance of red colours as expected from the removal of satellites.  
Therefore our global conclusion of the large-scale modulation of colours is not affected
by the modeling of galaxy colours, but rather by the underlying DM halo mass function which consistently
shows a larger population of high mass haloes further away from void centres, and closer to
cluster centres.

Our second test is related to a possible issue that may affect the observational results from 
\cite{cec}.  In their work they
intend to perform a comparison between galaxies at different distances from void centres, but within
equal local density environments.  To this aim, they construct a low local
density sample by applying a maximum density cut.  
This approach is adequate for sparse samples such as can be extracted for these studies, however
it is possible that the mean or median local densities in these subsamples are different for
different large-scale environments.  This bias can be ruled out
in the case of voids by examining Figure \ref{fig:fig2b}, since the distribution functions of IVD at different
distances from the void centres (right panels) show very similar shapes.  On the other hand, this would be important in
the study of galaxy colours at different distances from halo centres, since the shape of the density
distribution shows important changes as the distance to massive haloes increases.
Figure \ref{fig:figiso} shows the effect of using upper limits for low local density subsamples
(solid line), which shows a stronger dependence of the red fractions with the distance to massive DM haloes than would be
obtained using fixed median values (dashed lines), with differences of similar amplitude to that found between 
our samples at different LS environment with respect to their random counterparts.

\subsection{Dependence of the mass of galaxy host haloes with the environment}

\begin{figure*}
\begin{minipage}{180mm}
  \centering
  \vspace{0pt}
  \includegraphics[angle=0,width=0.6\columnwidth]{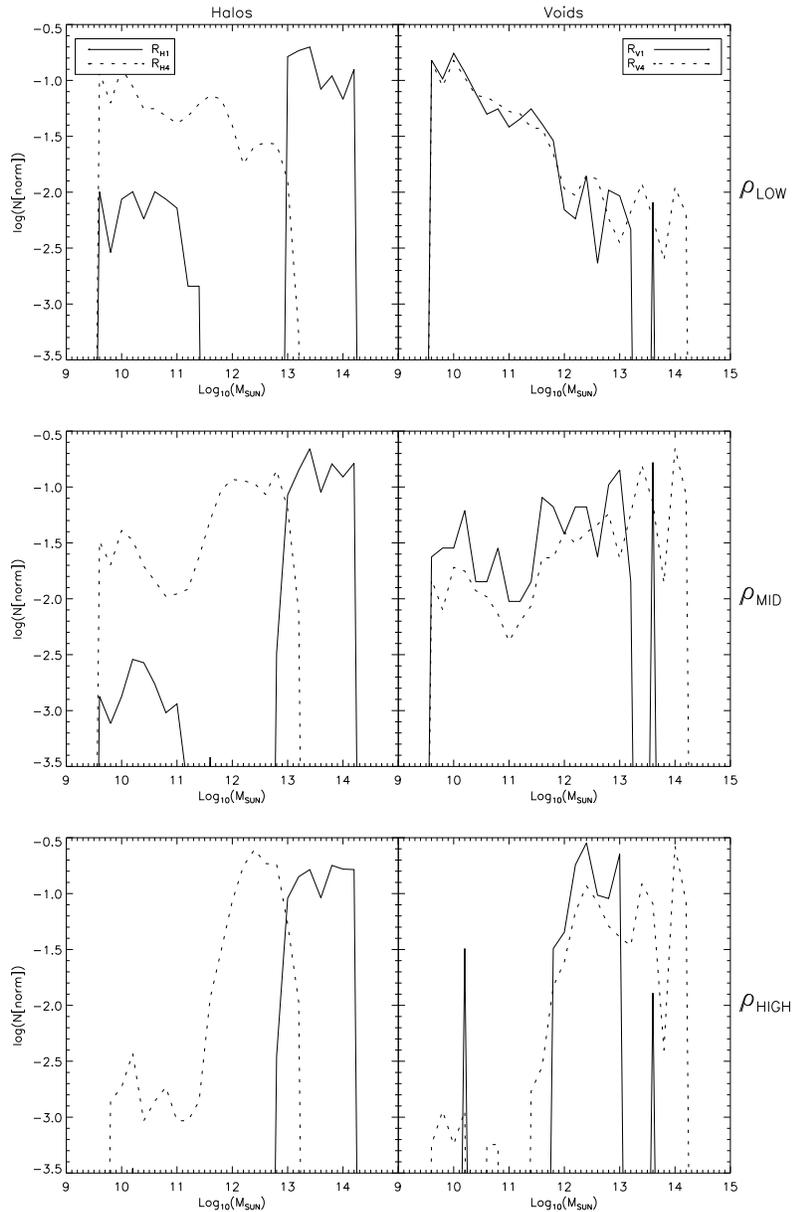}
  \caption{DM Host halo mass functions for different halo (left) and void (right) centric
distances and different local density cuts (increasing from top to bottom, indicated on the right-hand
axes of the figure).  The solid lines correspond to samples inside voids and haloes; the
dashed lines to samples at large distances from voids and haloes.}
  \label{fig:fig5}
\end{minipage}
\end{figure*}

The distribution functions of galaxy host halo masses change with both the local and global density environments
(Figure \ref{fig:fig5}).  The right panels of the figure show the mass functions for samples $R_{V1}$ and $R_{V4}$ (solid
and dashed lines, respectively) for increasingly higher local density environments from the top
to the bottom panels.
The left panels correspond to samples at different distances from massive haloes,
$R_{H1}$ and $R_{H4}$. Notice that these distributions cannot be compared to mass functions since
several galaxies are hosted by the same halo, and therefore many haloes are counted several times.
As the distance to massive haloes diminishes there is a clear increase in the typical host halo mass.  Also,
at a fixed distance, higher local densities are characterised by more massive host haloes.  It is
noticeable that this behaviour is also present at different distances from voids; 
the number of high mass host haloes diminishes both, towards the void centres and lower local density regions.

\subsection{Fractions of star-forming galaxies as a function of the environment}

\begin{figure}
  \vspace{0pt}
  \includegraphics[angle=0,width=0.9\columnwidth]{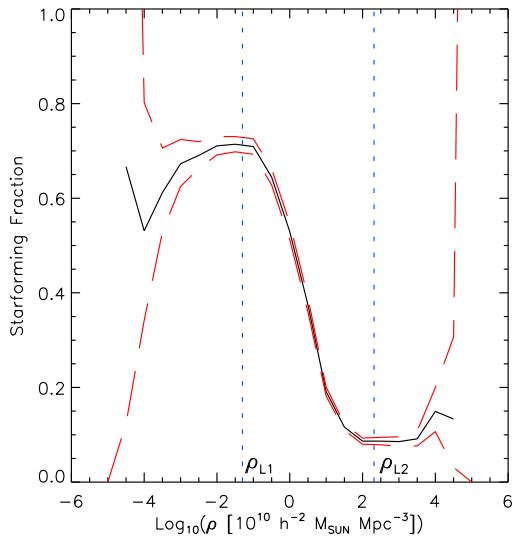}
  \caption{Dependence of the SF fraction on local density (solid lines).  The dashed lines 
show the amplitude of the poisson variance.  The vertical dotted lines indicate the densities
where the behaviour of the SF fraction shows marked changes.}
  \label{fig:fig7}
\end{figure}

Since the simulation contains information on the instantaneous SF of galaxies, we can also define a 
star-forming galaxy fraction for which we will adopt a lower limit on the SF rate of
$0.014 M_{\odot} yr^{-1}$, a value which corresponds to the median of the SF distribution at the $z=0$ simulation output.
In observations, the SF can be obtained via measurements of line strengths in the spectra
of galaxies.  Figure \ref{fig:fig7}, 
shows the dependence of the star-forming fraction as a function of the IVD, which shows a clear trend of a lower
frequency of star-forming galaxies towards high densities ($8$ times lower than in 
the lowest density environments), in agreement with the dependence of
$W_0(H_{\alpha})$ with the measured projected surface density $\Sigma_5$  in the SDSS\citep{bal}, and also with
results on the dependence of the red galaxy fraction with $\Sigma_5$.  In our results,
we find that even the lowest density environments contain some passive, non-star-forming galaxies which might
be associated to fossil groups \citep{do}; furthermore, for
$\rho_{L1}< 0.05$h$^3 Mpc^{-3}$ the fraction of star-forming galaxies remains constant, a result also
noticed by \citet{wei}; this provides a 
handle on the distance scale where galaxy-galaxy interactions may start to
become important (as was mentioned above,  in the case of \small{SAG}, this would correspond to halo-galaxy
interactions).  As the IVD increases, the star-forming fraction decreases steeply to
$\rho_{L2} \approx f_b \times 200 \rho_C$, where there is another break in the measured trend. 
Such density values occur at $\simeq r_{200}$, which is similar to the virial radius of the cluster, 
a distance at which galaxy properties may 
simply depend on the cluster environment rather than on interactions with neighbors.
Regarding this result, \citet{gom} found that at projected local densities corresponding to cluster-centric
distances of $1$ to $2$ virial radii,
the observed SF rate decreases rapidly and converges to a SF value which shows almost no variation
within the cluster virial radius, consistent with our results.
 
\begin{figure*}
\begin{minipage}{180mm}
  \centering
  \vspace{0pt}
  \includegraphics[angle=0,width=0.8\columnwidth]{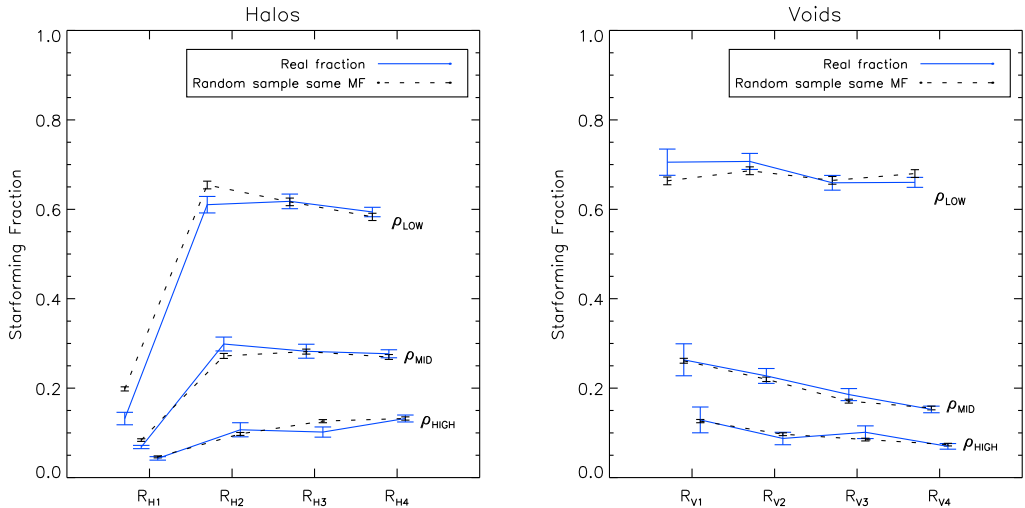}
  \caption{Star-forming fraction for different halo/void distances and different local density cuts.  Panels
and line styles are as in Fig. 6.}
  \label{fig:fig6}
\end{minipage}
\end{figure*}

We also study the variation of the fraction
of star-forming galaxies with the local and global density estimates.  We perform this study
using the same samples as in the previous sections, and show the results in 
Figure \ref{fig:fig6}.
Our findings are largely consistent with our results for red galaxy fractions; namely,
(i) towards halo centres the fraction of star-forming galaxies decreases (solid lines),  
(ii) for $R_{H2}$ at high local densities, specially at $\rho_{MID}$, we find a higher 
star-forming fraction($2.14\sigma$ significance) than in the random samples, in agreement with \citet{por}, and (iii)
 for $R_{H2}$ 
at low local densities there are less star-forming galaxies than in the random counterpart($2.16\sigma$ significance).
A slight
difference with respect to the red fractions is seen 
for the samples $R_{H1}$ at $\rho_{LOW}$ 
and $\rho_{MID}$  
where the fraction of star-forming galaxies is slightly lower than in the random counterparts.
These slight discrepancies between red and star-forming fraction trends can be explained 
by considering that galaxy colours integrate the formation history of galaxies, and therefore
take longer periods of time to migrate to blue colours.  For instance, considering that
the trends in star-forming and red fractions are compatible for samples $R_{V1}$ to $R_{V4}$ (low local densities)
and indicate a larger fraction of red/passive galaxies away from voids, while their corresponding random samples show
a larger fraction of red galaxies but a constant fraction of star-forming galaxies,
the delayed colour transformations would need to affect the counterpart random
samples more strongly in this case. 
As will be shown in Section 5.2 these random samples are characterised by slightly younger ages than their parent
samples, a result in favour of this picture.

The general conclusion from this analysis also supports that a global, large-scale effect is acting on the 
star formation  activity beyond the effects of the host halo mass function and small-scale interactions.
The SF seems to be strongly inhibited at the virial radius, but at larger halo distances it quickly converges to standard 
fractions which depend slightly on the large-scale environment.
In voids it can {be} seen that at low local densities there is a change in the star-forming fractions
from $R_{V1}$ to $R_{V2}$ with respect to their random counterpart(although with low significances, $1.36\sigma$ and $1.05\sigma$, respectively), 
in qualitative agreement with \citet{cec}.  This dependence tends to disappear at distances larger than the void radii,
particularly for the high local density regions.

In haloes, the star-forming fractions in the $R_{H1}$ samples differ quite significantly from fractions at larger halo distances.
This result is consistent with those of \citet{gom} who find that the fraction of star-forming galaxies in the SDSS decreases rapidly 
once the projected local galaxy number density increases  above a critical value of $\approx 1 $h$_{75}^{-2}$Mpc$^{-2}$ (see also Lewis et al., 2002). 
This density corresponds to the virial radii approximately, and therefore this measurement is interpreted as evidence
of strong changes in the morphological fractions due to different processes taking place within clusters or groups
of galaxies, such as the stripping of the warm gas in the outer halo of the infalling galaxies via tidal interactions
with the cluster potential.
This process has been shown to remove the hydrogen from the cold disk of the galaxy, and thus slowly 
strangle its star formation activity. 
Possible observational candidates for galaxies going through this process are red or passive spirals with negligible star 
formation, found at high \citep{couch} and low redshifts \citep{van2}, known as "anemic spirals".

\section{What causes a large-scale modulation of SF?}

In the previous section we demonstrated that the SF process in a $\Lambda$CDM cosmology shows variations
that cannot be accounted for by changes in the local density alone.  In particular, we have shown that
large-scale modulations in the fractions of red galaxies are mostly due to variations in the distributions
of host halo masses (as suggested by \citealt{cec}).  
This answers the question for the cause of the bulk of the dependence found in the SF of
galaxies.  However, there are still small but significant differences between low local density galaxies 
located in different large-scale environments
and their random/same mass function counterparts;
particularly for $R_{H1}$($4.51\sigma$), $R_{H2}$($2.16\sigma$), $R_{V1}$($1.36\sigma$) and $R_{V2}$($1.05\sigma$) at $\rho_{LOW}$, and $R_{H2}$($2.14\sigma$) at $\rho_{MID}$.  

There are different possible scenarios to explain the additional changes in galaxy colours to those
coming from the mass function.  In particular, the literature
proposes two possibilities, one related to coherent dynamics of the environments of galaxies
and the other to the age of the galaxy population; it should be noticed that it is likely
that these two quantities are correlated.

\subsection{Coherent motions}

\citet{por} propose that the enhancement of SF at $\approx 2.5 h^{-1} Mpc$ from the centres of haloes may be due to 
close interactions between galaxies falling into clusters from filaments; in this case the interactions would need to
occur before their gas reservoirs are stripped off by the intracluster medium.
In addition, \citet{cec} propose that the enhancement of star formation in void walls may be due to interactions
with galaxies arriving from the void inner regions (in addition to a possible effect from the local halo mass function).  
Both ideas propose an effect related to the coherence of galaxy
flows from or towards particular regions in the large-scale structure, and are particularly suitable for tests using
numerical simulations which we study next.

\begin{figure*}
\begin{minipage}{180mm}
  \centering
  \vspace{0pt}
  \includegraphics[angle=0,width=0.8\columnwidth]{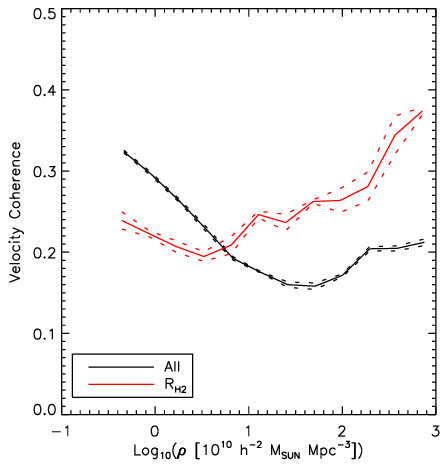}
  \caption{Velocity coherence $C$ as a function of local IVD. The solid black line corresponds
to the full \small{SAG} galaxy population; the red solid line corresponds to the $R_{H2}$ subsample.  Dashed lines
show the errors of the mean.}
  \label{fig:fig1co}
\end{minipage}
\end{figure*}

We compute the local velocity coherence for galaxies in all our subsamples at different local and global densities,
using the individual vector velocities of their $N_j$ voronoi neighbors,
\begin{eqnarray}
C_j = \frac{1}{N_j} \sum_{i=1}^{N_j} \frac{\vec{V_j}  \vec{V_i}}{|V_j||V_i|}.
\end{eqnarray}
This quantity provides information about the net flow of a galaxy and its neighbors; $C=1$ indicates that
galaxies are moving in the same direction; in the case of isotropic velocities, $C=0$.
Fig. \ref{fig:fig1co}
shows $C_j$ as a function of local density for all galaxies (black solid lines, dashed lines indicate the 
error of the mean). 
As can be seen, there is a significant trend of a decreasing coherence as the local density increases,
up to $\rho\simeq10^{1.5}$h$^{-2}M_{\odot}$Mpc$^{-3}$.

Table \ref{table:table1} shows the average velocity coherence for selected subsamples corresponding to 
(i) galaxies within the virial radius of haloes (first three rows, for different local densities), 
(ii) galaxies just outside the virial radius (rows four to six), (iii) galaxies in voids (penultimate row), and (iv)
galaxies in void walls (bottom row). 
The third column in the table shows the
$C$ values obtained for the samples of random galaxies selected to have the same distribution of host halo masses and
IVD densities; this is
shown in order to distinguish effects coming from parameters other than the local density and halo masses
(these samples are only a representative selection useful for further discussion).
These results illustrate that inside haloes
($R_{H1}$ in all local densities) the coherence is disrupted as expected within virialised regions.

For galaxies lying just outside the virial radius (from $1.5$ to $4\times r_{200}$, corresponding to $R_{H2}$ subsample) 
we find two different behaviours.  The lower local density subsamples show similar results to the $R_{H1}$
samples, with lower coherences than found in the corresponding random samples 
possibly indicating galaxies in eccentric orbits around the halo centres.
However, the sample corresponding to high
local density regions, $\rho_{HIGH}$, shows an enhanced coherence with respect to the corresponding random subsample
indicating coherent motions towards halo centres (already reported by e.g. Pivato, Padilla \& Lambas, 2006). 
This can also be seen in a black solid line in Figure \ref{fig:fig1co}, where the coherence 
increases towards higher local densities for galaxies located within this range of distances to halo centres.
This particular result can be related to the scenario proposed by \citet{por}, 
by noticing that
the local densities $\rho_{MID}$ and $\rho_{HIGH}$, 
just outside the virial radii of haloes, can be associated to filaments.
In our results we find an enhanced coherence within these regions as well as a slightly higher SF;
at high local densities the SF fraction is only $0.67\sigma$ higher than 
in the random counterpart, but at intermediate local densities this signal increases to a $2.14\sigma$ significance.
These results are in agreement with those presented by  Porter et al. (2008).

For different distances to void centres, $R_{V1}$ and $R_{V2}$, in the low local density $\rho_{LOW}$ case, 
galaxies show enhanced SF in comparison with the samples of random galaxies with the same distributions of
host halo masses($1.36\sigma$ for $R_{V1}$ and $1.05\sigma$ for $R_{V2}$), but only marginally 
larger coherences.  Therefore it is only hinted, but not possible to confirm from this analysis, that 
galaxies arrive at void walls in a slightly more ordered fashion than would correspond to galaxies
with similar underlying
distributions of host-halo masses and local densities, as suggested by \citet{cec}.

\begin{table}
\centering
Velocity Coherence\\
\begin{tabular}{@{}lcc@{}}
\hline
Subsample & $C_{SAMPLE}$ & $C_{RANDOM}$ \\
\hline
$R_{H1}$ $\rho_{LOW}$  & 0.0805 $\pm$ 0.0199  &  0.1708 $\pm$ 0.0023 \\
$R_{H1}$ $\rho_{MID}$  & 0.0747 $\pm$ 0.0056  &  0.1192 $\pm$ 0.0087 \\
$R_{H1}$ $\rho_{HIGH}$ & 0.1575 $\pm$ 0.0216  &  0.1980 $\pm$ 0.0048 \\
$R_{H2}$ $\rho_{LOW}$  & 0.2232 $\pm$ 0.0117  &  0.3085 $\pm$ 0.0031 \\
$R_{H2}$ $\rho_{MID}$  & 0.2157 $\pm$ 0.0075  &  0.2536 $\pm$ 0.0024 \\ 
$R_{H2}$ $\rho_{HIGH}$ & 0.3438 $\pm$ 0.0049  &  0.2943 $\pm$ 0.0162 \\
$R_{V1}$ $\rho_{LOW}$  & 0.3584 $\pm$ 0.0056  &  0.3511 $\pm$ 0.0021 \\
$R_{V2}$ $\rho_{LOW}$  & 0.3571 $\pm$ 0.0031  &  0.3502 $\pm$ 0.0018 \\
\hline
\end{tabular}
\caption{Velocity Coherence for the selected subsamples indicated in the first column. $C_{SAMPLE}$ \& $C_{RANDOM}$ correspond to the mean coherence shown by the sample and its corresponding random counterpart with same local density and mass functions.}
\label{table:table1}
\end{table}

\subsection{Age dependence on the large-scale distribution of matter}

\begin{figure*}
\begin{minipage}{180mm}
  \centering
  \vspace{0pt}
  \includegraphics[angle=0,width=0.8\columnwidth]{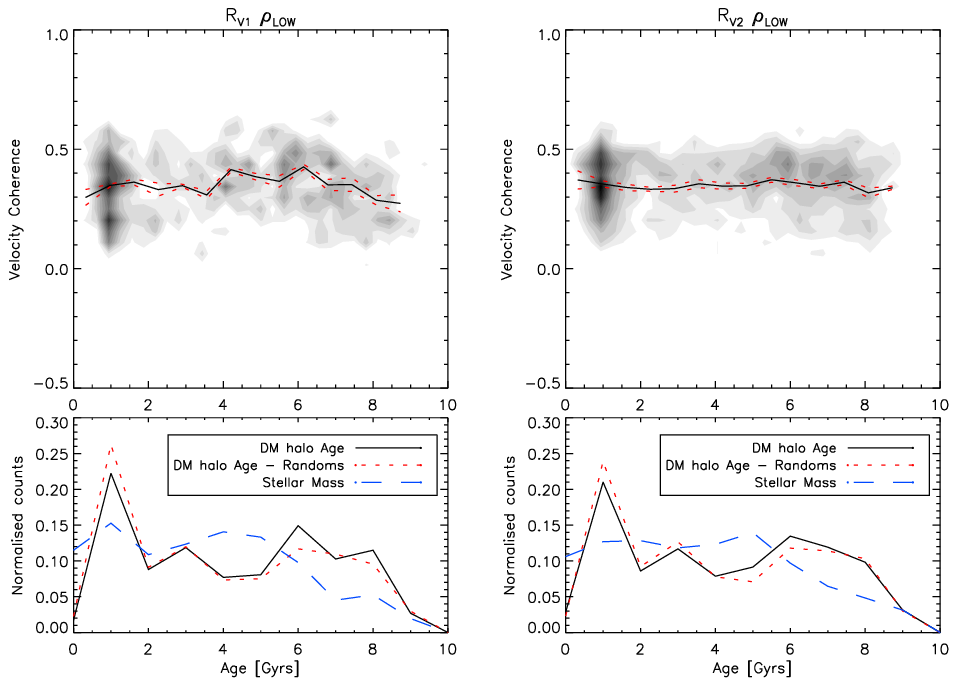}
  \caption{Velocity coherence as a function of Age.  The different shades of grey indicate the density of DM haloes with a given 
coherence and age (defined as $z_{1/2,t_1}$; darker colours indicate higher density). The left panels
show the results for sample $R_{V1}$, $\rho_{LOW}$.  The right panels, for $R_{V2}$ $\rho_{LOW}$. 
The bottom panels show the normalised distributions of DM halo age ($z_{1/2,t_1}$, solid lines), DM halo ages for the 
associated random samples (dashed lines),
and stellar mass weighted ages (long-dashed lines).}
  \label{fig:fig2co}
\end{minipage}
\end{figure*}

Another galaxy parameter that has recently gained attention in the field is the galaxy age, or the halo
assembly epoch (e.g. Zapata et al., 2009).  The samples with fixed local density and varying
large-scale environments are indeed characterised by different stellar mass-weighted ages than
their random counterparts (with equal distribution of host halo mass and median local density); in particular, galaxies
in lower global densities and low local density environments tend to be older
than their random counterparts when using stellar ages 
($0.61\sigma$ for $R_{V1}$ and $\rho_{LOW}$) 
an effect which becomes more noticeable when using DM halo ages
obtained from $z_{1/2,t_1}$ defined as in Li, Mo \& Gao (2008) 
($1.14\sigma$ effect).  Stellar and halo ages are indistinguishable between random and parent samples
for higher local densities.
Furthermore, the mass-weighted stellar
ages show different degrees of correlation with the formation time of the DM halo at different
global environments.
Fig. \ref{fig:fig2co} depicts
the coherence as a function of DM halo age 
(darker colours indicate a higher density of galaxies in the age vs. coherence plane) 
for the two subsamples $R_{V1}$ 
and $R_{V2}$ for low local densities, $\rho_{LOW}$.  The
solid line shows the average coherence, with its corresponding error bars shown in red dashed lines. 
The bottom panels show the normalised distributions of DM halo assembly
and stellar mass weighted ages corresponding to the samples shown
in the upper panels (black solid and red dotted lines, respectively), and 
DM halo assembly ages for the random counterpart samples (long-dashed lines).

As can be seen, the DM halo assembly ages in both subsamples 
tend to be slightly older than in their corresponding random subsamples ($1.14\sigma$ for $R_{V1}$, and 
$0.76\sigma$ for $R_{V2}$).  Haloes residing in void walls 
do not show significant differences in their ages with respect to haloes inside voids; however,
the stellar populations in void walls do show a slightly clearer tendency towards older ages.

Finally, when the samples are restricted to fixed local and global densities, the age vs. coherence
correlation is apparently completely erased (at least there is no significant detection of a correlation).  This would
indicate that the origin of the age-coherence correlation observed for the full
sample of galaxies is mostly due to an underlying DM halo mass vs. coherence relation.

\section{Conclusions}

In this paper we have studied large-scale modulations of galaxy colours and SF rates at different fixed local
density environments, within the framework of a $\Lambda$CDM model.  
In order to parameterise a global large-scale environment we used distances to landmarks given by
massive haloes and large-scale voids in the simulation.  Also, given the advantage of the knowledge of the
underlying properties of galaxies and their host DM haloes, we were able to study the cause
for such modulations.

We now summarise the main results of this study.
\begin{itemize}
\item The large-scale structure produces subtle but noticeable effects on the SF rate of galaxies beyond the 
local density dependence.  We find that
the global
density has an important impact in colours and SF at low local densities, whereas local galaxy-galaxy interactions may 
be the major responsibles for the variations seen in galaxy colours and SF rates at higher local density environments, 
where galaxy-galaxy interactions become important.
\item In voids we find a large-scale influence beyond the local density dependence in agreement with 
\citet{cec}.  This dependence weakens at distances larger than a few void radii.
\item Inside haloes, the SF of galaxies is strongly suppressed possibly due to the 
gas depletion processes implemented in the model when galaxies are acquired as new satellites.
\item Galaxies just outside the virial radius in high density regions (filaments) show indications
of infalling motions towards clusters.  Additionally, these galaxies show traces of enhanced SF activity at intermediate
densities, $\rho_{MID}$.  Also, for intermediate and high local density environments, galaxies just outside haloes show
an enhanced SF activity associated to a higher velocity coherence, in agreement with the results by \citet{por}.
\item  The largest differences in SF fractions  with respect to their corresponding 
random subsamples (selected to reproduce the MF and IVDs, constructed using
galaxies in any global environment) are seen for galaxies 
apparently approaching void walls from the inner void regions.  A possible cause
for this effect could come from the velocity coherence which is seen
to differ with respect to the associated random sample (although with a low statistical significance; see Figure
\ref{fig:fig6}). 
\item We find that the connection between coherence and age found for the full sample of \small{SAG} galaxies
weakens dramatically when restricting the global and local environments simultaneously, leaving
only slight differences in halo ages.  This indicates that the age-coherence relation is a result of
an underlying correlation between age and the host DM halo mass function.
\item 
Taking into account this last result,
assembly ages seem better suited as candidates to produce the differences in galaxy colours and SF found between
the parent and random samples.
\end{itemize}
Our main conclusion is that the large-scale environment affects the galaxy population mostly via
variations in the mass function.  However, there appears to be a slight correlation between
assembly and large-scale structure, an effect that could be tested using large
observational datasets providing a new, innovative
way to place further constraints on our current paradigm of structure and galaxy formation.
However, the present status of observational results grant another important success
to the $\Lambda$CDM model which as we showed in this work, is able to reproduce unexpected aspects of the
variations of galaxy properties with the environment.

\section*{Acknowledgments}
We thank Michael Drinkwater for very useful comments which significantly improved this paper.
This work was supported in part by the FONDAP ``Centro de Astrof\'\i sica" and Fundaci\'on
Andes.  NP was supported by a Proyecto Fondecyt Regular No. 1071006.

\end{document}